\documentclass[11pt]{article}

\usepackage{acl}

\usepackage{times}
\usepackage{latexsym}

\usepackage[T1]{fontenc}

\usepackage[utf8]{inputenc}

\usepackage{microtype}

\usepackage{inconsolata}

\usepackage{graphicx}
\usepackage{amsmath}
\usepackage{multirow}
\usepackage{amssymb}
\usepackage{booktabs}
\usepackage{subfig}

\title{AgentTrace: Execution-Level Fingerprinting of Black-Box LLM Agents}

\author{
	Gejian Zhao, Hanzhou Wu, and Xinpeng Zhang \\
	Shanghai University \\
	\texttt{\{23820171, hanzhou, xzhang\}@shu.edu.cn}
}

\begin{document}
\maketitle
\begin{abstract}
	Verifying the provenance of autonomous LLM agents is a critical security challenge. Existing semantic attribution is easily obfuscated by system prompts, while active watermarking conflicts with schema-constrained tool generation. We propose \textbf{AgentTrace}, an active framework that identifies the foundation model behind a black-box agent via observable execution traces. AgentTrace injects structurally ambiguous probes to trigger tokenization-sensitive serialization failures at rigid tool interfaces. While these initial failures provide coarse-grained separation across model families, the temporal sequence of subsequent repair operations reveals fine-grained, model-specific recovery behaviors. By modeling these multi-step trajectories with lightweight Markov chains, AgentTrace achieves 96.0\% attribution accuracy under adversarial identity spoofing. Crucially, sequential repair modeling successfully distinguishes the evaluated intra-family variants sharing identical tokenizers (e.g., Llama-3 8B/70B), substantially outperforming unordered behavioral statistics while remaining robust against prompt obfuscation and temperature scaling.
\end{abstract}

\section{Introduction}
Large language models (LLMs) are increasingly deployed as autonomous agents that plan, reason, and execute tasks through tool integration \cite{li2026deepagent,dwivedi2026artificial}. These agents are often exposed as opaque black-box services, with foundation models hidden behind APIs, system prompts, and orchestration layers \cite{10748304}. This layered design makes provenance verification substantially harder than in standard conversational settings \cite{he2025emerged}. As such agents increasingly interact with critical infrastructure, verifying their provenance becomes not only an ownership concern but also a security requirement.

A pressing threat in this opaque ecosystem is \textit{model spoofing}, where providers covertly wrap high-performance models under false proprietary claims \cite{gubritrap, pasquini2025llmmap}. Unmasking such agents exposes fundamental limitations in current fingerprinting techniques. Passive methods relying on $n$-gram distributions \cite{tayetal2020reverse}, perplexity \cite{mitchell2023detectgpt}, or stylometry \cite{uchendu2020authorship} fail because agentic prompts effectively rewrite outputs, destroying raw statistical assumptions \cite{gudibande2024false}. Furthermore, active watermarking \cite{kirchenbauer2023watermark} introduces a severe utility-security trade-off in structured generation (e.g., JSON), as probability perturbations disrupt the strict syntactic precision required by external tools \cite{ajith2024downstream,11495247}.
\begin{figure*}[t]
	\centering
	\includegraphics[width=0.87\linewidth]{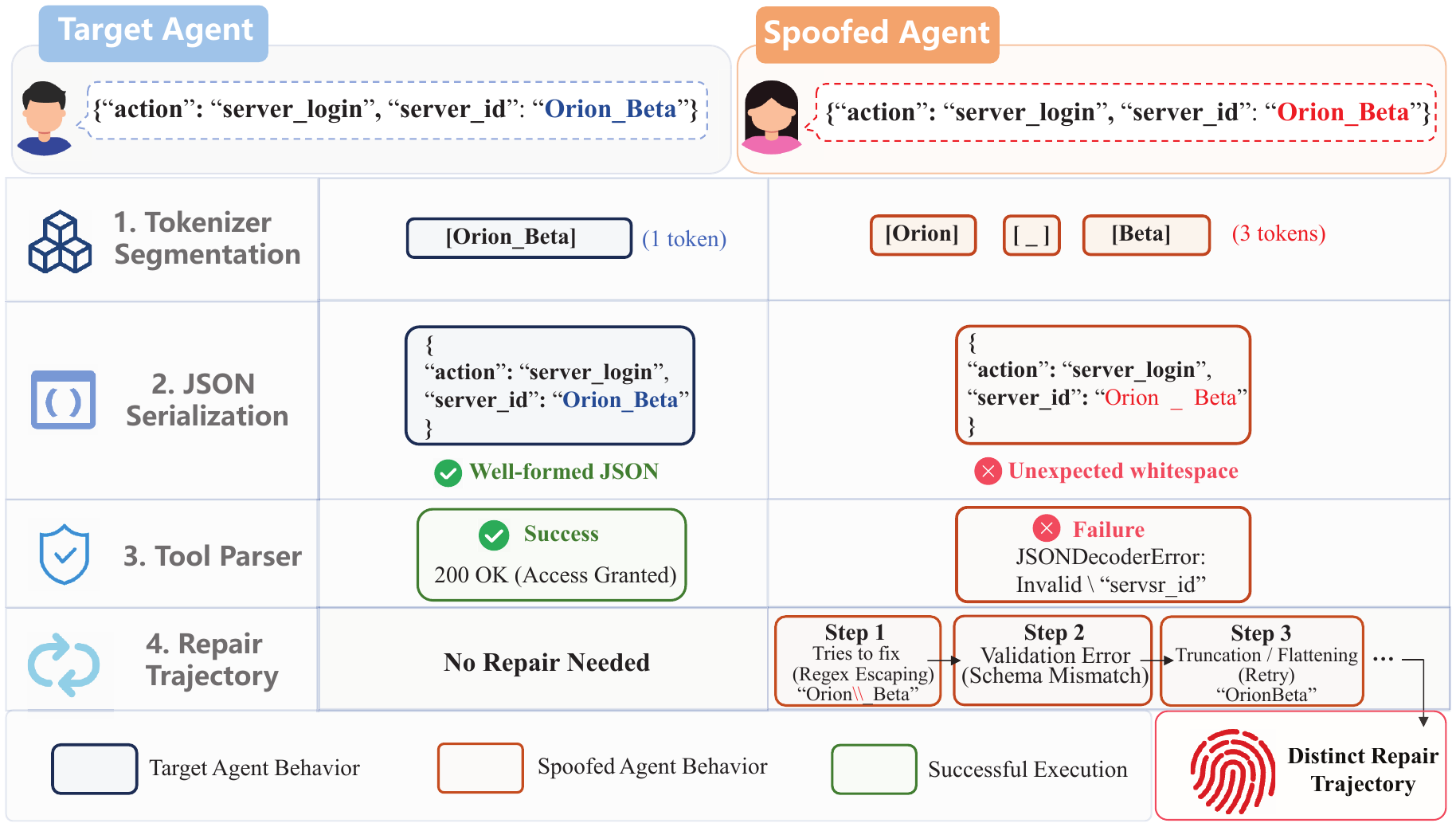} 
\caption{Conceptual workflow of AgentTrace. A genuine target model may reproduce a structurally ambiguous identifier successfully. A spoofed model with a different tokenization pattern may produce malformed escaping or other serialization anomalies rejected by the parser. AgentTrace combines the tokenization-sensitive failure category with the subsequent sequence of repair operations to identify the underlying model.}
	\label{fig:teaser}
\end{figure*}

To break this impasse, we pivot from semantic fingerprinting to execution profiling (Figure~\ref{fig:teaser}). Consider an agent tasked with generating a schema-constrained API identifier. While a target model might tokenize this string efficiently, a spoofed agent relying on a different tokenizer may partition it into disparate subwords. We empirically observe that such tokenization discrepancies often induce reconstruction instability during serialization, manifesting as subtle structural artifacts like spurious whitespace. Although semantically benign in natural language, these artifacts are deterministically amplified into execution failures by rigid tool parsers. 

By treating attribution as an active sequential identification task, we propose \textbf{AgentTrace}. Unlike semantic probing methods that query textual boundaries, AgentTrace searches for adversarial probes maximizing the discriminative information within multi-step execution traces. By systematically inducing and profiling these traces, our framework unmasks the underlying model. Crucially, AgentTrace utilizes a two-tiered attribution signal: the initial tokenization-sensitive failure provides coarse-grained separation, while the subsequent model-specific repair policy exposes fine-grained behavioral signatures that remain empirically stable across our evaluated settings, transcending prompt obfuscation \cite{singh2024tokenization}.

In summary, our main contributions are:
\begin{itemize}
	\item We identify a novel two-tiered execution-level attribution signal: tokenization-sensitive execution outcomes for cross-family separation, and model-specific sequential repair policies for intra-family differentiation.
	\item We propose AgentTrace, an active framework that uses mutual information as an intuitive motivation to guide system identification, utilizing multi-dimensional repair trajectories to unmask spoofed agents.
	\item We construct a comprehensive benchmark and demonstrate state-of-the-art attribution performance. Our ablation studies confirm that modeling sequential repair behaviors is essential for separating identically-tokenized variants.
\end{itemize}

\section{Preliminaries}
\label{sec:preliminaries}

\subsection{Problem Formulation}
\label{subsec:problem_formulation}

\paragraph{Black-Box Agent and Execution Trace} 
Formally, a tool-augmented LLM agent is defined as $\mathcal{A} = \langle M^*, \mathcal{P}, \mathcal{T}, \mathcal{O} \rangle$, where $M^*$ is the underlying foundation model, $\mathcal{P}$ contains system instructions, $\mathcal{T}$ represents external tools, and $\mathcal{O}$ is the orchestration mechanism (including parsers, schema validators, and retry logic). 

Assuming private chain-of-thought is hidden, an input probe $s$ yields an externally observable trace $\tau_s =[(a_1, o_1), \dots, (a_L, o_L)]$. At step $t$, $a_t$ is the serialized tool argument, and $o_t = (b_t, e_t)$ is the deterministic observation, where $b_t \in \{\text{Success}, \text{Error}\}$ is the execution status and $e_t$ denotes the exception category (e.g., \texttt{JSONDecodeError}).

\paragraph{The Attribution Task}
Let $\mathcal{M} = \{M_1, \dots, M_N\}$ be a known candidate set. Given a collection of probe-induced traces $\mathcal{Q} = \{\tau_{s_1}, \dots, \tau_{s_Q}\}$, the auditor's goal is to infer the hidden foundation model from the aggregated traces:
\begin{equation}
	\hat{M} = F(\mathcal{Q}), \quad \hat{M} \in \mathcal{M}.
\end{equation}

\subsection{Threat Model}
\label{subsec:threat_model}
We operate under a strict black-box setting, which is highly representative of real-world endpoint-substitution attacks:
\begin{itemize}
	\item \textbf{Auditor Capabilities:} The auditor can query the agent and observe tool arguments, parser feedback, and final outcomes. They lack access to weights, logits, or private prompts, but possess local/API access to reference models in $\mathcal{M}$.
	\item \textbf{Adversary Capabilities:} The adversary covertly replaces $M^*$ with a substitute, and may modify system prompts or decoding parameters (e.g., temperature) to conceal the identity.
	\item \textbf{Orchestration Scope:} Crucially, the client-side orchestration $\mathcal{O}$, tool schemas $\mathcal{T}$, parser, and retry logic remain fixed. This models scenarios where an application's infrastructure is intact while backend requests are maliciously redirected.
\end{itemize}

\section{Methodology}
\label{sec:methodology}
AgentTrace operates through a unified three-stage pipeline: (I) Offline Probe Optimization, (II) Online Execution Injection, and (III) Sequential Trace Differentiation. 

\subsection{Theoretical Foundation}
\label{subsec:theoretical_insight}
Conceptually, we frame active attribution as discovering an optimal probe $s^*$ that maximizes the mutual information between the candidate model identity $M \in \mathcal{M}$ and the induced trace $\tau_s$:
\begin{equation}
	s^* = \arg\max_{s} I(M ; \tau_s)
	\label{eq:mi}
\end{equation}
Directly estimating Equation~\ref{eq:mi} is intractable in a black-box setting. Therefore, AgentTrace optimizes a practical surrogate based on the empirical execution-failure probability. Let $\mathcal{G}_M(\cdot \mid s, \mathcal{P})$ denote the autoregressive distribution over a serialized output $y$. We define the execution gap as the probability of a structural or schema violation:
\begin{equation}
	\begin{aligned}
		\mathrm{Gap}(M,s)
		= \Pr_{y\sim\mathcal{G}_M}\big[
		&\mathrm{Parse}(y)=\mathrm{Error} \\
		&\lor\ \mathrm{Valid}_{\mathcal{T}}(y)=\mathrm{Error}
		\big].
	\end{aligned}
	\label{eq:gap}
\end{equation}
Rigid parsers transform latent generation instability into categorical outcomes, bridging the heuristic gap to our conceptual goal without requiring access to logits.

\subsection{Stage I: Offline Probe Optimization}
\label{subsec:stage1}
To operationalize the attribution, AgentTrace constructs a diverse offline library of structurally ambiguous probes. For an input string $s$, let $B_M(s)$ denote the set of byte offsets where tokenizer $\mathrm{Tok}_M$ places boundaries. The normalized token-boundary distance between models $M_i$ and $M_j$ is:
\begin{equation}
	D_{\text{tok}}(M_i, M_j; s) = 1 - \frac{|B_{M_i}(s) \cap B_{M_j}(s)|}{|B_{M_i}(s) \cup B_{M_j}(s)|}
\end{equation}
To ensure robust multi-class attribution, the offline search optimizes a joint surrogate score $J(s)$ that balances target-verification margin ($\Delta_{\text{target}}$), execution-gap variance across all candidates ($\Delta_{\text{exec}}$), and pairwise tokenizer diversity ($\Delta_{\text{tok}}$):
\begin{equation}
	J(s) = \Delta_{\text{target}}(s) + \lambda_e \Delta_{\text{exec}}(s) + \lambda_t \Delta_{\text{tok}}(s)
\end{equation}
Specifically, let $M_c$ denote the claimed target and
$\mathcal{M}_{-c}=\mathcal{M}\setminus\{M_c\}$ the substitutes.
We define
\begin{equation}
	\Delta_{\text{target}}(s)
	=
	\overline{\mathrm{Gap}}_{\mathcal{M}_{-c}}(s)
	-\mathrm{Gap}(M_c,s),
\end{equation}
where $\overline{\mathrm{Gap}}_{\mathcal{M}_{-c}}$ is the mean
failure probability over substitutes. We further set
$\Delta_{\text{exec}}$ to the variance of $\mathrm{Gap}(M,s)$
over $\mathcal{M}$ and $\Delta_{\text{tok}}$ to the mean pairwise
token-boundary distance.

The search enforces $\mathrm{Gap}(M_c,s)\le\epsilon$ and an
average substitute failure rate of at least $\gamma$, and is
optimized by mutation-based beam search.

\subsection{Stage II: Online Execution Injection}
\label{subsec:stage2}
During auditing, AgentTrace sequentially injects selected probes into the black-box agent via a structured tool argument. A representative instruction is:
\begin{quote}
	\small
	\textit{``Execute the function \texttt{[Tool\_Name]}. Set the field \texttt{[Argument\_Name]} exactly to the following value: \texttt{[Probe]}. Preserve every character and do not summarize.''}
\end{quote}
This template forces the model to reproduce the probe at the boundary of language generation and deterministic software execution. If the initial call fails, the orchestration framework appends the error feedback, prompting the model to generate a corrected call until success or a retry limit is reached.

\subsection{Stage III: Sequential Trace Differentiation}
\label{subsec:stage3}

Given a trace $\tau_s=\{(a_1,o_1),\dots,(a_L,o_L)\}$, we represent its
behavior as $z_\tau=\langle E,\mathcal{R}_{\text{seq}}\rangle$, where
$E$ is the initial outcome and $\mathcal{R}_{\text{seq}}$ is the ordered
repair sequence. We factorize
$P(z_\tau\mid M)=P(E\mid M)P(\mathcal{R}_{\text{seq}}\mid E,M)$.
For $L=1$, the repair sequence is empty and
$P(\mathcal{R}_{\text{seq}}\mid E,M)=1$. For $L>1$,
\begin{equation}
	\begin{aligned}
		P(\mathcal{R}_{\text{seq}}\mid E,M)
		=&\;P(r_1\mid E,M)\\
		&\times\prod_{t=2}^{L-1}
		P(r_t\mid r_{t-1},E,M).
	\end{aligned}
\end{equation}
The probabilities are estimated from offline profiling with
add-one Laplace smoothing.

Given $Q$ observed traces, AgentTrace predicts
\begin{equation}
	\hat{M}=\arg\max_{M\in\mathcal{M}}
	\sum_{q=1}^{Q}
	\frac{\log P(z_{\tau_{s_q}}\mid M)}
	{\max(1,L_q-1)}.
	\label{eq:attribution}
\end{equation}
The normalization prevents long retry trajectories from dominating the attribution score.

\begin{table}[t]
	\centering
	\small
	\caption{Evaluation models and tokenizer configurations.}
	\label{tab:models}
	\resizebox{\columnwidth}{!}{%
		\begin{tabular}{@{}lllc@{}}
			\toprule
			\textbf{Role} & \textbf{Family} & \textbf{Model Instance} & \textbf{Tokenizer} \\ \midrule
			Claimed Target & GPT & \texttt{gpt-4o} & o200k\_base \\ \midrule
			\multirow{5}{*}{Substitutes} & \multirow{2}{*}{Llama} & \texttt{Llama-3-8B-Instruct} & \multirow{2}{*}{Tiktoken BPE, 128k} \\
			& & \texttt{Llama-3-70B-Instruct} & \\ \cmidrule(l){2-4} 
			& Mistral & \texttt{Mistral-7B-Instruct-v0.3} & SentencePiece, 32k \\ \cmidrule(l){2-4} 
			& \multirow{2}{*}{Qwen} & \texttt{Qwen-2.5-7B-Instruct} & \multirow{2}{*}{Tiktoken BPE, 152k} \\
			& & \texttt{Qwen-2.5-72B-Instruct} & \\ \bottomrule
		\end{tabular}%
	}
\end{table}

\section{Experiments and Analysis}
\label{sec:evaluation}
In this section, we evaluate the effectiveness, robustness, and granularity of AgentTrace. Our experiments are designed to answer the following research questions:
\begin{itemize}
	\item \textbf{RQ1 (Attribution Accuracy):} Can AgentTrace accurately unmask the underlying foundation model of a black-box agent, even under strict adversarial spoofing?
	\item \textbf{RQ2 (Granularity):} Can sequential trace differentiation distinguish between evaluated models of the same family sharing identical tokenizers (e.g., Llama-3 8B/70B)?
	\item \textbf{RQ3 (Robustness):} Is the framework resilient against defensive system prompts and varied generation temperatures?
\end{itemize}

\subsection{Setup}
\label{subsec:setup}
\paragraph{Target Models: Claimed Identity vs. Surrogates}
We select a representative set of LLMs covering distinct tokenizer families and varying parameter scales. To simulate the unauthorized wrapping ecosystem, we categorize them into two roles:
\begin{itemize}
	\item \textbf{The Target Identity (Ground Truth):} We use GPT-4o (OpenAI) as the impersonated target, representing a flagship proprietary model frequently targeted for spoofing.
	\item \textbf{The Surrogate Models (Spoofers):} We select open-weight model families utilized as covert engines: (1) \textbf{Llama-3 (8B \& 70B)} and \textbf{Qwen-2.5 (7B \& 72B)}, which allow us to rigorously evaluate intra-family granularity on identically-tokenized variants across different scales; and (2) \textbf{Mistral-7B-Instruct-v0.3}, featuring a distinct tokenizer vocabulary.
\end{itemize}
All open-weight surrogates are deployed locally via \texttt{vLLM} with the generation temperature set to $T=0.1$ (unless testing robustness). GPT-4o is accessed via the official API.

\paragraph{Agent Environment}
All candidate models are integrated into the same LangChain-based ReAct orchestration stack equipped with a strict \texttt{JSONAgentOutputParser}. The agents interact with three rigid external tools: a typed JSON API, a regex-constrained file-system interface, and a Python code executor. These tools impose strict physical boundaries: any malformed serialization immediately triggers a deterministic parser exception, which is appended to the context window for subsequent recovery attempts.

\paragraph{Simulated Spoofing Scenario}
To rigorously evaluate performance under active obfuscation, each substitute model is injected with a deceptive system prompt instructing it to claim that it is GPT-4o and strictly conceal its true identity. This configuration fundamentally breaks traditional text-based authorship attribution by explicitly overwriting the agent's semantic identity.

\paragraph{Probe Library and Data Split}
Using the offline search detailed in Stage I, we construct a library of $K=100$ distinct adversarial probes. To rigorously prevent data leakage, we strictly partition the library into $K_{\text{train}} = 50$ profiling probes (used exclusively to estimate the Markov transition matrices) and $K_{\text{test}} = 50$ held-out probes for final testing. All test traces are generated independently from the profiling traces.

\paragraph{Baselines and Metrics}
We compare AgentTrace against two semantic baselines:
\begin{itemize}
	\item \textbf{Stylometry:} A traditional classifier based on lexical and $n$-gram features.
	\item \textbf{Cross-Perplexity:} Attributes outputs to the candidate assigning the lowest normalized perplexity.
\end{itemize}
To validate our specific framework design, we also evaluate two ablations: \textbf{Initial Outcome} (uses only the first parser result) and \textbf{Bag-of-Operations} (tallies repair actions but discards their temporal sequence). 

\textit{Note on recent active baselines:} While recent active fingerprinting methods like LLMMap~\cite{pasquini2025llmmap} achieve high accuracy, they are designed for conversational text interfaces and do not directly define an attribution protocol over schema-constrained execution traces. A faithful adaptation would require fundamentally changing their observable interface and probing procedure, making direct comparison non-equivalent. Therefore, they are not included in our agentic baselines.

\paragraph{Evaluation Protocol}
To compute robust attribution metrics and clarify the unit of evaluation, we evaluate performance at a query budget of $Q$ probes. For the main results in Table 2, we set $Q=50$. The effect of varying $Q$ is evaluated separately. For each candidate model, we perform 100 independent evaluation trials. In each trial, we randomly sample $Q$ probes from the held-out test set without replacement and perform attribution via Equation~\ref{eq:attribution}. The variance across trials is driven by the stochastic generation and repair sampling of the models. The reported accuracy, macro-F1, and recall are averaged across these 100 trials.

\subsection{Main Results: Unmasking Spoofed Agents (RQ1)}
\label{subsec:rq1_main}

\begin{table*}[t!]
	\centering
	\small
	\caption{Attribution performance on the held-out probes under adversarial identity spoofing.}
	\label{tab:main_results}
	\begin{tabular}{l|cc|cc|cc|cc}
		\toprule
		\multirow{2}{*}{\textbf{Underlying Model}} & \multicolumn{2}{c|}{\textbf{Stylometry}} & \multicolumn{2}{c|}{\textbf{Cross-Perplexity}} & \multicolumn{2}{c|}{\textbf{Bag-of-Operations}} & \multicolumn{2}{c}{\textbf{AgentTrace (Ours)}} \\
		\cmidrule{2-9}
		& \textbf{Recall} & \textbf{F1} & \textbf{Recall} & \textbf{F1} & \textbf{Recall} & \textbf{F1} & \textbf{Recall} & \textbf{F1} \\
		\midrule
		GPT-4o (\textit{Target Identity}) & 94.0\% & 0.48 & 88.0\% & 0.62 & 92.0\% & 0.85 & \textbf{98.0\%} & \textbf{0.99} \\
		\midrule
		Llama-3-8B-Instruct     & 18.0\% & 0.15 & 42.0\% & 0.38 & 72.0\% & 0.68 & \textbf{92.0\%} & \textbf{0.93} \\
		Llama-3-70B-Instruct    & 24.0\% & 0.20 & 46.0\% & 0.42 & 74.0\% & 0.70 & \textbf{94.0\%} & \textbf{0.95} \\
		Mistral-7B-Instruct     & 12.0\% & 0.11 & 38.0\% & 0.35 & 78.0\% & 0.78 & \textbf{98.0\%} & \textbf{0.98} \\
		Qwen-2.5-7B-Instruct    & 14.0\% & 0.12 & 48.0\% & 0.40 & 70.0\% & 0.65 & \textbf{96.0\%} & \textbf{0.96} \\ 
		Qwen-2.5-72B-Instruct   & 16.0\% & 0.13 & 54.0\% & 0.48 & 76.0\% & 0.76 & \textbf{98.0\%} & \textbf{0.98} \\
		\midrule
		\textbf{Overall Accuracy} & \multicolumn{2}{c|}{29.6\%} & \multicolumn{2}{c|}{52.6\%} & \multicolumn{2}{c|}{77.0\%} & \multicolumn{2}{c}{\textbf{96.0\%}} \\
		\textbf{Macro-F1}         & \multicolumn{2}{c|}{0.19} & \multicolumn{2}{c|}{0.44} & \multicolumn{2}{c|}{0.73} & \multicolumn{2}{c}{\textbf{0.96}} \\
		\bottomrule
	\end{tabular}
\end{table*}

To answer RQ1, we evaluate the attribution accuracy of AgentTrace against the semantic baselines under the active spoofing scenario ($\mathcal{P}_{spoof}$). The overall performance is summarized in Table~\ref{tab:main_results}.

\paragraph{Vulnerability of Semantic Baselines}
Our experiments empirically validate that traditional attribution methods degrade significantly in agentic workflows. Under the deceptive prompt, the substitute models effectively adjust their visible stylistic outputs to mimic GPT-4o. Consequently, both stylometry and cross-perplexity exhibit severe False Acceptance Rates. This confirms that semantic text features are easily overwritten by system instructions, rendering them unreliable for auditing.

\paragraph{Robust Attribution via Execution Probing}
In contrast, AgentTrace achieves 96.0\% overall accuracy and 0.96 macro-F1, successfully unmasking the underlying models despite intense identity obfuscation. By bypassing the mutable semantic layer entirely, AgentTrace effectively combines the coarse-grained signal of initial tokenizer-sensitive crashes with the fine-grained signal of ordered repair trajectories. Furthermore, as demonstrated by the gap between Bag-of-Operations (77.0\%) and AgentTrace (96.0\%), explicitly modeling the sequential temporal order of recovery transitions provides a decisively stronger fingerprint than merely tallying unordered repair actions.

\paragraph{Mechanism Validation: Tokenization Sensitivity}
To empirically validate our core hypothesis that tokenization differences influence observable parser failures, we analyze the relationship between tokenizer boundaries and execution outcomes. For our library of optimized probes, a Spearman correlation analysis reveals a significant positive correlation ($\rho = 0.54$, $p < 0.01$) between the token-boundary distance $D_{\text{tok}}$ and the empirical execution gap difference between cross-family models. This provides direct empirical evidence that structural parsing anomalies are indeed sensitive to underlying tokenization mechanics. Furthermore, ablating the $\Delta_{\text{tok}}$ term from the probe optimization function degrades the overall attribution accuracy to 84.5\%, confirming that explicitly optimizing for token-level divergence is crucial for discovering high-quality probes.

\begin{table}[t]
	\centering
	\small
	\caption{Effect of the query budget ($Q$) on attribution accuracy.}
	\label{tab:query_budget}
	\resizebox{0.9\columnwidth}{!}{%
		\begin{tabular}{@{}lccccc@{}}
			\toprule
			\textbf{Query Budget ($Q$)} & \textbf{1} & \textbf{3} & \textbf{5} & \textbf{10} & \textbf{50} \\ \midrule
			\textbf{Accuracy (\%)} & 65.2 & 78.4 & 86.5 & 92.8 & 96.0 \\ \bottomrule
		\end{tabular}%
	}
\end{table}

\paragraph{Query Efficiency and Practicality}
To evaluate how attribution performance scales with API cost, we vary the query budget $Q \in \{1, 3, 5, 10, 50\}$. As shown in Table~\ref{tab:query_budget}, attribution accuracy improves rapidly as the query budget increases. Because our Markov MLE formulation effectively aggregates log-likelihood evidence across multiple independent repairs, it resolves stochastic generation noise early in the sampling process. Notably, AgentTrace achieves 92.8\% accuracy with only $Q=10$ probes. This demonstrates that reliable model attribution can be achieved with a highly practical and low-cost API query budget.

\subsection{Fine-Grained Attribution via Repair Trajectories (RQ2)}
\label{subsec:rq2_granularity}

A key limitation of purely tokenization-based fingerprints is their inability to distinguish intra-family variants sharing identical vocabularies. For example, Llama-3-8B and Llama-3-70B adopt the same tokenizer and therefore exhibit nearly identical first-order tokenization behaviors. Similarly, Qwen-2.5-7B and 72B share an identical vocabulary. Consequently, our adversarial probes frequently induce the exact same initial execution failure ($E_{type}$) across variants of the same family, making the tokenizer-execution gap alone insufficient for fine-grained attribution.

To answer RQ2, we evaluate whether the sequential repair process following the initial failure provides additional discriminative information. Although same-family variants exhibit similar crash patterns, their subsequent recovery trajectories differ substantially due to differences in model capacity and generation behavior.

\begin{figure}[t]
	\centering
	\includegraphics[width=0.99\linewidth]{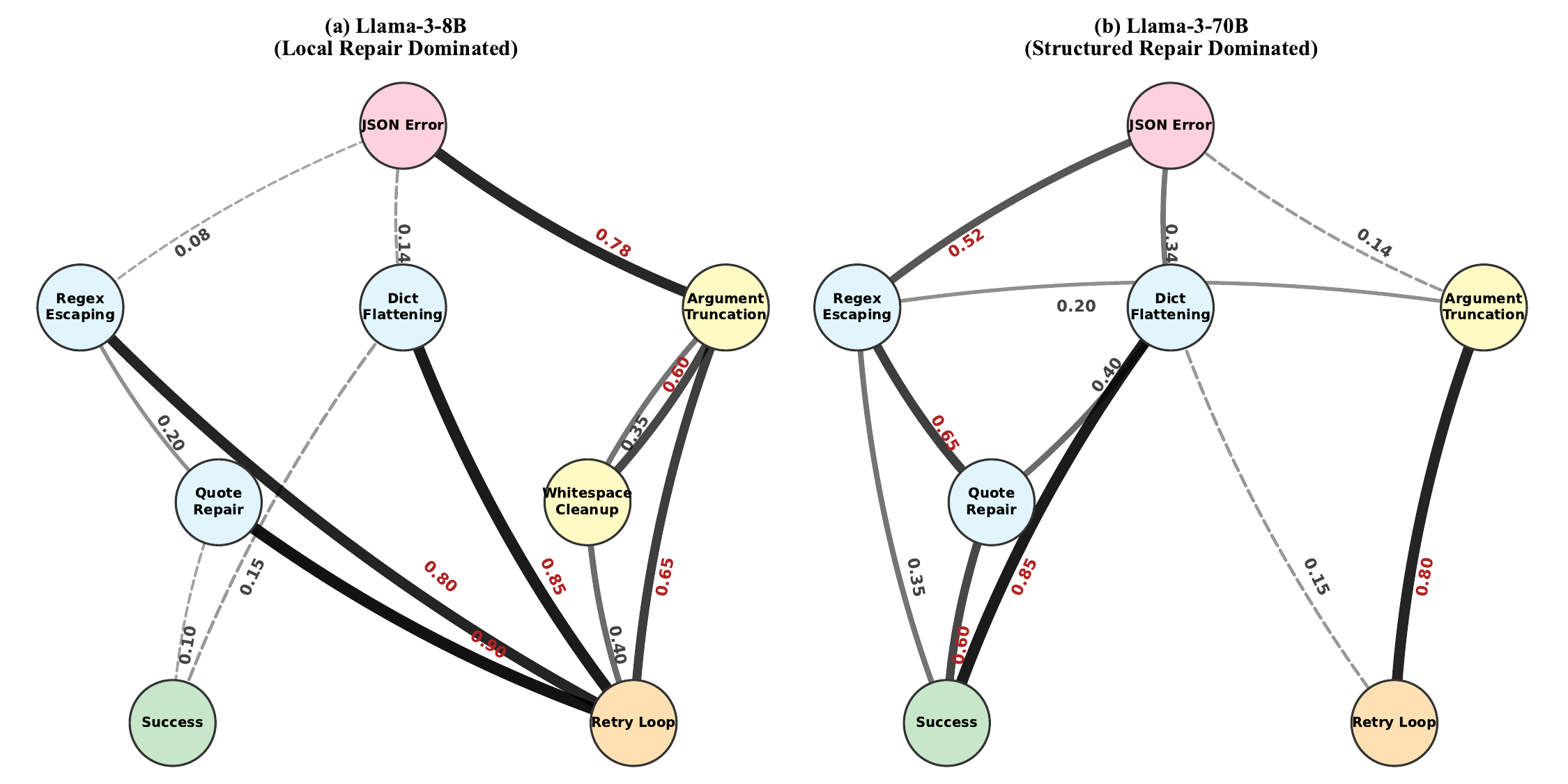}
	\caption{Empirical repair-transition graphs for Llama-3 variants (edge thickness corresponds to transition probability). While both models suffer identical initial tokenization crashes, their repair strategies strictly diverge: (a) Llama-3-8B is trapped in repetitive correction loops, while (b) Llama-3-70B navigates toward success via structured repair.}
	\label{fig:trajectory_markov}
\end{figure}

Figure~\ref{fig:trajectory_markov} visualizes empirical transition graphs derived from the induced repair traces. We observe that highly capable models (e.g., Llama-3-70B) frequently apply structured operations (e.g., regex escaping) to achieve successful execution. Conversely, smaller counterparts (e.g., Llama-3-8B) rely on local modifications (e.g., argument truncation), often trapping themselves in retry cycles. These divergent policies allow AgentTrace's Markov modeling to capture behavioral signals completely invisible at the tokenizer level.

\begin{table}[t]
	\centering
	\caption{Ablation study on execution-trace representations.}
	\label{tab:ablation}
	\resizebox{0.98\columnwidth}{!}{%
		\begin{tabular}{@{}lccc@{}}
			\toprule
			\textbf{Representation} & \textbf{Llama-3 Pair} & \textbf{Qwen-2.5 Pair} & \textbf{Overall} \\ \midrule
			Initial Outcome & 50.0\% & 50.0\% & 66.8\% \\
			\textbf{+ Repair Trajectory} & \textbf{93.0\%} & \textbf{97.0\%} & \textbf{96.0\%} \\ \bottomrule
		\end{tabular}%
	}
\end{table}

To quantify the necessity of this sequential modeling, Table~\ref{tab:ablation} presents our ablation study. Relying solely on the initial crash indicator (Initial Outcome) yields only 66.8\% overall accuracy. Crucially, for intra-family variants with identical tokenizers, performance collapses to a 50.0\% random guess. Integrating the Markovian repair trajectory restores intra-family accuracy to 93.0\% for Llama-3 variants and 97.0\% for Qwen-2.5 variants, confirming that multi-step recovery dynamics provide the essential discriminative signal for fine-grained attribution.

\subsection{Robustness Against Defensive Interventions (RQ3)}
\label{subsec:rq3_robustness}

\begin{figure}[t]
	\centering
	\subfloat[Prompt Obfuscation]{
		\includegraphics[width=0.47\linewidth]{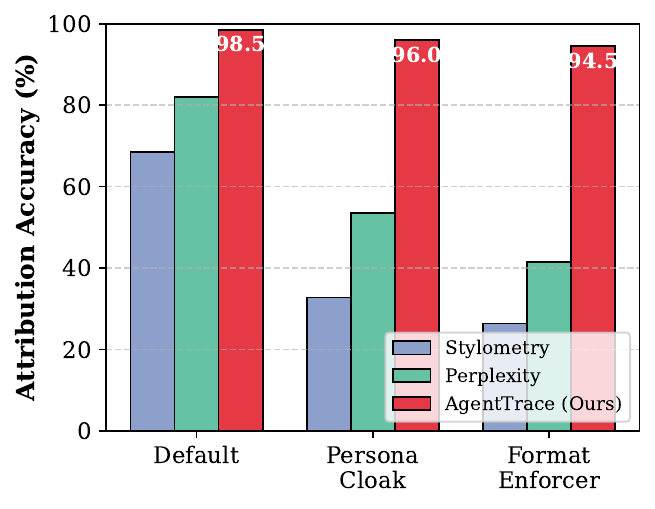}
	}
	\hfill
	\subfloat[Temperature Scaling]{
		\includegraphics[width=0.47\linewidth]{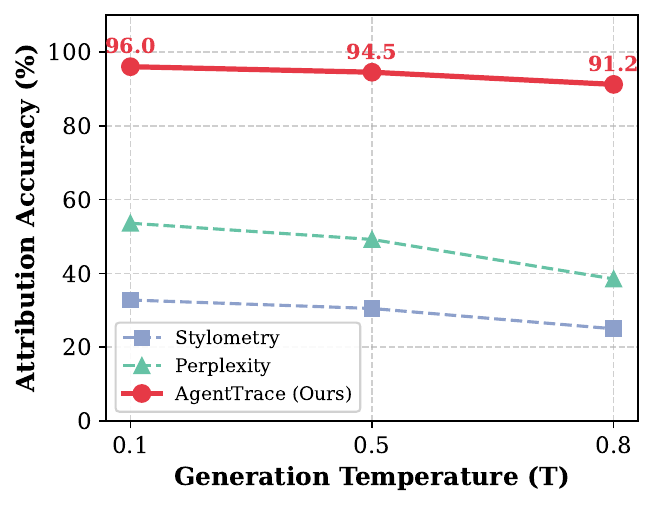}
	}
	\caption{Attribution accuracy under defensive interventions. (a) Semantic baselines degrade substantially under prompt-based obfuscation, whereas AgentTrace remains stable across different system prompts. (b) Although higher generation temperatures introduce stochasticity, the Markovian trajectory modeling in AgentTrace maintains robust attribution performance.}
	\label{fig:robustness}
\end{figure}

In real-world model spoofing scenarios, adversaries may actively employ defensive interventions to evade attribution. To answer RQ3, we evaluate AgentTrace against two primary evasion strategies: system prompt obfuscation and generation temperature scaling.

\paragraph{Resilience to Prompt Obfuscation}
We assess the framework under three distinct system prompt configurations: \textbf{Default} (standard ReAct agent), \textbf{Persona-Cloak} (explicitly denying the true identity), and \textbf{Format-Enforcer} (strictly mandating raw JSON outputs without extraneous characters). As shown in Figure~\ref{fig:robustness}(a), semantic baselines degrade drastically under both \textit{Persona-Cloak} and \textit{Format-Enforcer}, as the modified instructions heavily rewrite the textual output distributions. 

In contrast, AgentTrace maintains over 94\% attribution accuracy across all prompt settings. This empirical stability confirms our theoretical intuition: while prompt engineering operates at the semantic alignment layer, it cannot alter the hardcoded tokenization mechanics required to invoke tools. To bypass the agent framework, the model must reproduce the adversarial probe at the rigid tool interface.

\paragraph{Resilience to Temperature Scaling}
Adversaries may elevate generation temperatures (e.g., $T \in \{0.1, 0.5, 0.8\}$) to introduce stochasticity, which increases repair variance and complicates exact trace matching. However, rather than relying on strict sequence matching, AgentTrace computes the joint log-likelihood of multi-step traces via its Markov transition model. Although individual repair actions become noisier at higher temperatures, the underlying transition probabilities remain highly discriminative. As shown in Figure~\ref{fig:robustness}(b), our Maximum Likelihood Estimation (MLE) effectively absorbs this generation noise, maintaining robust attribution accuracy even at $T=0.8$.

\subsection{Qualitative Analysis and Case Study}
\label{subsec:case_study}

\begin{figure}[t]
	\centering
	\includegraphics[width=0.95\linewidth]{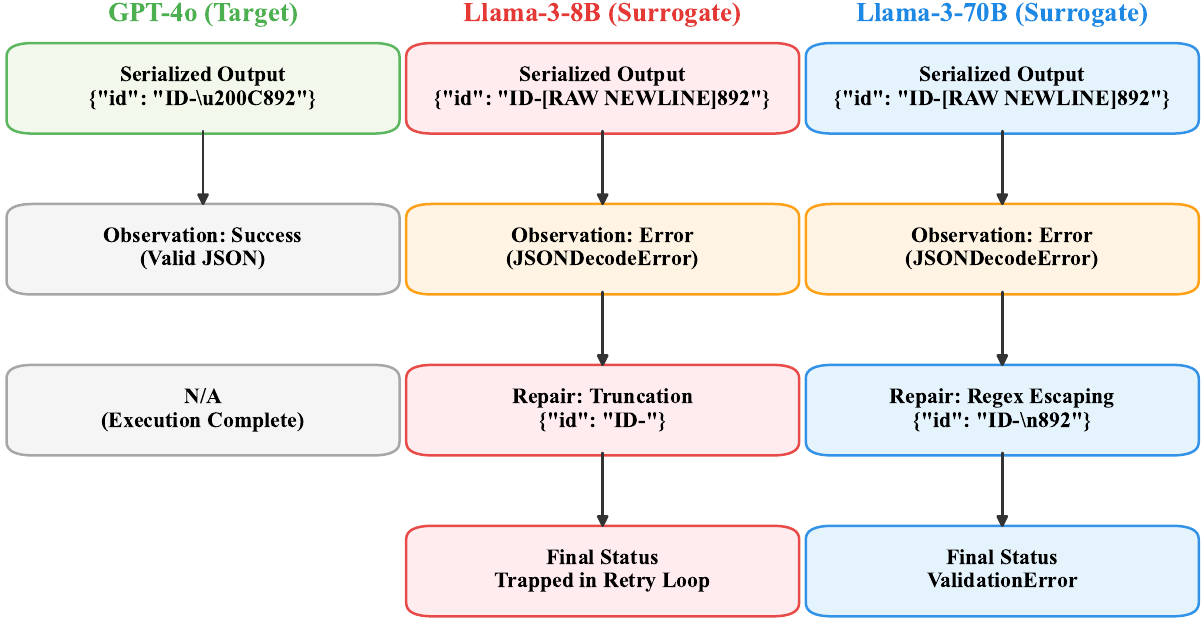} 
	\caption{Multi-step execution traces induced by the same adversarial identifier. Despite identical initial JSON errors, the Llama-3 variants exhibit divergent repair dynamics: the 8B model falls into truncation retry loops, whereas the 70B model employs structured escaping. These distinct sequential trajectories provide robust execution-level fingerprints.}
	\label{fig:case_study_trace}
\end{figure}

To ground our probabilistic attribution in observable system behaviors, we present a concrete case study illustrating the physical tokenization mismatch and the subsequent Markovian repair trajectory. 

Figure~\ref{fig:case_study_trace} visualizes a representative multi-step execution trace induced by an adversarial identifier containing a zero-width non-joiner (\texttt{U+200C}): \texttt{"ID-\textsubscript{ZWNJ}892"}.

\paragraph{Tokenization Discrepancy}
The target model, GPT-4o, typically preserves this original Unicode boundary. Instead of consistently triggering serialization failures, it produces syntactically valid JSON payloads with high probability, fulfilling the schema constraints and executing successfully.

Conversely, the surrogate Llama-3 variants fragment the string into atypical subword segments. We empirically observe that this atypical fragmentation increases the instability of constrained serialization. During autoregressive JSON generation, this perturbation frequently induces the model to output a raw, unescaped newline character directly after the ZWNJ. 

\paragraph{Sequential Repair Trajectory}
The raw newline causes a \texttt{JSONDecodeError}, after which the two Llama-3 variants exhibit distinct repair behaviors. \textbf{Llama-3-8B} frequently truncates the argument, leading to a secondary \texttt{ValidationError} and repeated retry failures. In contrast, \textbf{Llama-3-70B} more often applies structured JSON escaping (e.g., generating \texttt{"\textbackslash n"}) and proceeds toward successful recovery.

This divergence, such as the transition \texttt{JSONDecodeError} $\to$ \texttt{Json-Escape}, illustrates why modeling the temporal order of repair operations provides a stronger fingerprint than static textual features. Similar repair divergences are observed across other structurally ambiguous probes, showing how rigid tool interfaces amplify tokenization-sensitive serialization differences into discriminative execution behaviors.

\section{Related Work}
\label{sec:related_work}
\subsection{LLM Fingerprinting and Watermarking}
Determining the provenance of LLM-generated content is a rapidly evolving field. Early works focused on passive fingerprinting, exploiting statistical artifacts such as $n$-gram frequencies \cite{tayetal2020reverse} or the local curvature of log-probability distributions, as exemplified by DetectGPT \cite{mitchell2023detectgpt}. While effective for base models, these methods rely on the assumption that the output reflects the raw distribution. This assumption is frequently violated in agentic workflows, where system prompts and persona conditioning act as high-pass semantic filters, effectively masking these statistical signals \cite{uchendu2020authorship, wang-etal-2023-seqxgpt}.

To mitigate these limitations, recent research has pivoted towards active watermarking, which embeds detectable patterns by perturbing token probabilities \cite{kirchenbauer2023watermark, zhao2024provable}. However, such perturbations often introduce a utility-security trade-off in agentic workflows \cite{lee2024wrote}: in structured domains like code or JSON generation, probability deviations can disrupt the strict syntactic constraints required by external tools. This renders traditional watermarking challenging to deploy reliably in task-oriented agents.

\subsection{Active Probing and Model Attribution}
Active probing methods attempt to infer hidden model properties through carefully designed interactions. Recent work such as LLMMap \cite{pasquini2025llmmap} formulates attribution as a semantic querying process, aiming to recover model decision boundaries from response behaviors. Other approaches similarly rely on eliciting distinctive textual outputs through adaptive prompts.

While promising, these methods primarily operate within the semantic space, making them sensitive to prompt engineering, alignment modifications, and output post-processing \cite{gudibande2024false,wang-etal-2023-seqxgpt}. As modern agents increasingly interleave generation with external tool execution, semantic outputs alone may no longer faithfully reflect the underlying model's inherent characteristics, necessitating signals beyond semantic boundaries.

\subsection{Tokenization and Agentic Vulnerabilities}
Tokenization constitutes the physical layer of LLMs, yet it is often overlooked in attribution analysis. Prior research has identified that tokenizer discrepancies can lead to performance drops in multilingual and code generation tasks \cite{singh2024tokenization, petrov2023language}. 

Security researchers have begun to exploit these discrepancies. For instance, ``glitch tokens'' have been shown to trigger irrational outputs \cite{zhang2024glitchprober}, and adversarial attacks such as ArtPrompt \cite{jiang2024artprompt} use atypical tokenization to bypass safety alignment. Similarly, \cite{geh2025adversarial} demonstrated that manipulating token boundaries can evade detection. While these works highlight the vulnerability of tokenizers to adversarial inputs, they primarily focus on safety evasion or jailbreaking. These findings suggest that tokenization artifacts may provide highly specific, model-level signals. Consequently, our work investigates whether such physical tokenization phenomena can be systematically exploited for robust model attribution in agentic environments.

\section{Conclusion}
\label{sec:conclusion}
In this paper, we introduced AgentTrace, an active execution-level framework for unmasking spoofed tool-augmented LLM agents. Rather than relying on mutable semantic outputs, AgentTrace shifts model attribution toward observable execution and recovery behaviors. By injecting structurally ambiguous probes into rigid tool interfaces, the framework exposes tokenization-sensitive serialization differences as externally observable execution outcomes. Beyond these coarse-grained failure patterns, AgentTrace further models the temporal structure of multi-step repair trajectories to capture fine-grained, model-specific recovery behaviors. Extensive experiments demonstrate that AgentTrace achieves strong attribution accuracy under adversarial identity spoofing, distinguishes the evaluated intra-family variants sharing identical tokenizers, and remains robust against prompt obfuscation and increased generation stochasticity. These results highlight that execution traces provide complementary provenance information that is difficult to capture through semantic fingerprinting alone. Overall, our findings suggest that execution and error-recovery dynamics offer a practical and promising signal for auditing the provenance of black-box agentic systems.

\section*{Limitations}
While AgentTrace demonstrates robust attribution capabilities, it relies on several operational assumptions. First, the framework requires the target agent to execute external tools with strict schema validation. Second, our sequential trace differentiation assumes the agent's middleware surfaces informative error feedback. Highly restrictive environments that forcibly terminate sessions after a single failure may limit the observable repair trajectory. 

Furthermore, AgentTrace fundamentally relies on imperfect serialization fidelity between the model and the tool interface. Adaptive middleware enforcing strict canonical serialization (e.g., applying \texttt{json.dumps()} or aggressively stripping invisible Unicode characters) may partially suppress tokenizer-induced artifacts. Finally, while our offline probe optimization yields high accuracy on the evaluated models, future work must investigate the transferability of these probes to unseen model checkpoints to mitigate potential overfitting to specific model quirks.

\section*{Ethics Statement}
This work aims to enhance the security, accountability, and intellectual property protection of LLM-based systems. By providing a robust auditing mechanism to detect unauthorized model wrapping, our framework empowers developers and organizations to enforce licensing compliance and verify system provenance. Although AgentTrace utilizes adversarial execution probes, the framework is intended solely for auditing and provenance verification in controlled evaluation settings. We acknowledge that attribution techniques may also be misused for unauthorized model surveillance or competitive auditing of deployed systems. Therefore, we advocate that such tools should only be used in legitimate security, compliance, or licensing verification scenarios to foster transparency within the evolving ecosystem of autonomous agents.

\bibliography{custom}

\end{document}